\documentclass[structabstract]{aa}  

\usepackage{graphicx}
\usepackage{enumerate}
\usepackage{txfonts}
\usepackage{natbib}
\usepackage{multirow}

\newcommand{\kepler}{{\sl Kepler }}
\newcommand{\keplers}{{\sl Kepler}'s }
\newcommand{\gaia}{{\sl Gaia }}
\newcommand{\gaias}{{\sl Gaia}'s }
\newcommand{\gaiasampled}{{\sl Gaia}-sampled }
\newcommand{\polyfit}{{\sl polyfit }}
\newcommand{\twog}{{\sl two-Gaussian }}

\begin{document}

   \title{\textit{Gaia} Eclipsing Binary and Multiple Systems. A study of detectability and classification of eclipsing binaries with \textit{Gaia}}

   \author{A.~Kochoska\inst{1,3}
          \and N.~Mowlavi\inst{2}
          \and A.~Pr\v sa\inst{3}
          \and I.~Lecoeur-Ta\"ibi\inst{4}
          \and B.~Holl\inst{2}
          \and L.~Rimoldini\inst{4}
          \and M.~S\"uveges \inst{4}
          \and L.~Eyer\inst{2}
          }

   \institute{
   University of Ljubljana, Deptartment of Physics, Jadranska 19, SI-1000 Ljubljana, Slovenia\\
   \email{angela.kochoska@fmf.uni-lj.si}
   \and
             University of Geneva, Department of Astronomy, 
             Chemin des Maillettes 51, CH-1290 Versoix, Switzerland
   \and
             Villanova University, Department of Astrophysics and Planetary Science, 800 Lancaster Ave, Villanova PA 19085, USA
   \and
            University of Geneva, Department of Astronomy, 
              Chemin d'Ecogia 16, CH-1290 Versoix, Switzerland\
}

   \date{Received 25 Oct 2016; accepted 27 Mar 2017}
 
  \abstract
   {In the new era of large-scale astronomical surveys, automated methods of analysis and classification of bulk data are a fundamental tool for fast and efficient production of deliverables. This becomes ever more imminent as we enter the \gaia era.} 
   {We investigate the potential detectability of eclipsing binaries with \gaia using a data set of all \kepler  eclipsing binaries sampled with \gaia cadence and folded with the \kepler period. The performance of fitting methods is evaluated with comparison to real \kepler data parameters and a classification scheme is proposed for the potentially detectable sources based on the geometry of the light curve fits.}
   {The polynomial chain ({\sl polyfit}) and \twog models are used for light curve fitting of the data set. 
   Classification is performed with a combination of the $t$-SNE ($t$-distrubuted Stochastic Neighbor Embedding) and DBSCAN (Density-Based Spatial Clustering of Applications with Noise) algorithms.}
   {We find that $\sim68\;\%$ of \kepler Eclipsing Binary sources are potentially detectable by \gaia when folded with the \kepler period and propose a classification scheme of the detectable sources based on the morphological type indicative of the light curve, with subclasses that reflect the properties of the fitted model (presence and visibility of eclipses, their width, depth, etc.).}
   {}


   \titlerunning{Gaia EBs - Detectability and classification}
   \authorrunning{A.~Kochoska et al.}
   \maketitle
%
\section{Introduction}

The onset of large-scale astronomical surveys is producing a steady flow of a large amount of data, which resulted in many ground-breaking discoveries made merely in the past few decades. Among the most important common objects that can be found in these surveys are binary stars, whose greatest contribution to astronomy is the possibility to directly measure stellar properties to an unprecedented level of accuracy.

With the use of a combination of observational techniques, in particular photometry and spectroscopy, we can obtain a full characterization of the system and its separate components: their orbital elements and dynamics, their absolute masses, radii, temperatures, chemical composition, rotation, the presence of other companions or planets etc. Thus, photometric-variability surveys such as \textit{Hipparcos}~\citep{perryman1997}, \textit{MOST}~\citep{pribulla2010}, \textit{CoRoT} \citep{corot}, \textit{OGLE-III}~\citep{udalski2008}, \textit{ASAS}~\citep{pojmanski2002} and \textit{\kepler}~\citep{keplermain}, have been of utmost importance to the field of binary stars, yielding extensive eclipsing binary star catalogs with data on several tens of thousands of stars. \gaia \citep{gaia2,2016A&A...595A...1G} is expected to boost this number by several orders of magnitude --- out of a billion observed sources, up to several million are expected to be eclipsing binaries (four million are predicted by \citealt{eyer2013}, seven million by \citealt{zwitter2002} and half a million by \citealt{dischler2005}). A portion of these (approximately $12\%$; \citealt{eyer2013}) are expected to be also spectroscopic binaries, which enables precise determination of their masses and radii.

The full physical characterization of spectroscopic eclipsing binary stars is a highly demanding and often incomplete task due to the parameter space degeneracy of the analysis models. A first step towards the characterization of an eclipsing binary is automated analysis of the geometrical parameters of its light curve (eclipse depth, width, separation, amplitude of ellipsoidal variations), which can be related to the physical system parameters such as periodicity, morphology, eccentricity, inclination, temperature ratio, etc. 
Likewise, the ever-growing data inflow calls for new, automated classification methods. Several machine learning methods, together with Principal Component Analysis \citep[PCA;][]{pca} and Locally Linear Embedding \citep[LLE;][]{lle}, $t$-Distributed Stochastic Neighbor Embedding \citep[$t$-SNE;][]{tsne} and Density-Based Spatial Clustering of Applications with Noise \citep[DBSCAN;][]{dbscan} have been considered and their performance evaluated on data from photometric and spectroscopic surveys \citep{caballero2008,kepler3,galrave,kepler7,suveges2016}.

In this paper, we propose a combination of the $t$-SNE and DBSCAN algorithms for the purposes of eclipsing binary light curve classification.

The methodology and results presented here are part of a series of exploratory studies undertaken in the framework of the \gaia mission for the implementation of an automated pipeline to process eclipsing binary light curves within the \gaia Data Processing and Analysis Consortium (eclipsing binary data from \gaia are expected to be delivered to the scientific community not earlier than 2019).
In this study, we use \kepler eclipsing binary light curves sampled with \gaia cadence at their respective positions on the sky.
We apply two different techniques to characterize the geometry of the folded light curves, and study the efficiency of the $t$-SNE and DBSCAN algorithms to classify the folded light curves.
As a by-product of this analysis we obtain an estimate of the \gaia recovery rate of Kepler eclipsing binaries, a number of interest to evaluate the eclipsing binary completeness factor expected from the \gaia mission.

An overview of the data sets, light curve fitting and classification methods is given in Sect.~\ref{Sect:Data_and_analysis}.
Results of the fitting models comparison are given in Sect.~\ref{Sect:Results}, as well as a classification of the whole and of a filtered data set of \gaiasampled light curves.
Conclusions of the paper and future prospects are given in Sect.~\ref{Sect:Conclusions}.


\section{Data and analysis}\label{Sect:Data_and_analysis}

\subsection{Overview of methodology}
\label{Sect:overview_of_methods}

We use data from the \kepler eclipsing binary catalog \citep{kepler7} sampled with the expected five year \gaia cadence to simulate \gaia light curves.
The \gaiasampled light curves are then folded using \kepler orbital periods and reference times of primary minimum (in barycentric Julian date), and fitted with two models.
The first model uses polynomial chain fits ({\sl polyfits}), as described in  \cite{ebaipaper}.
The second model, called the \twog model, chooses the best combination of Gaussian functions to describe the presence of eclipses and a cosine function to describe an ellipsoidal-like variability during the inter eclipses, if present. This model is developed within the \gaia pipeline to process the light curves of eclipsing binaries (Mowlavi et al., submitted).

The $t$-SNE algorithm requires a set of data computed in the same set of $x$-axis points. For this purpose, we fit the phase-folded data and compute all models in a set of $N$ equidistant phase points. As mentioned above, the orbital periods and reference times of primary minimum are fixed to their \kepler values. Further studies will rely on periods provided by the \gaia period-search pipeline. The \twog model fits are only folded with the orbital period, while the shifting to phase zero is done with respect to the phase at maximum magnitude of the phase-folded light curve.

Due to \keplers unprecedented photometric precision and high cadence observations of the original \kepler field of view in Cygnus, its light curves are of remarkable quality. \gaias scanning law, in contrast, observes a given star 67 times on average in a five year mission lifetime (the actual number of observations of a given star depends on its position in the sky and follows \gaias Nominal Scanning Law; see Sect.~\ref{Sect:gaiasampled}), which will result in insufficiently sampled light curves for a portion of the eclipsing binary sources. In our \gaiasampled set, the resulting light curves of these sources typically give poor or unrealistic model fits, which can be automatically isolated and removed by the fitting model itself, in the case of {\sl two-Gaussians}, or the above-mentioned dimensionality reduction and clustering algorithms, in the case of {\sl polyfits}. 

The dimensionality reduction and clustering method is then used to propose a classification scheme of the remaining sources.


\subsection{Data sets}
\label{Sect:datasets}

\subsubsection{Original \kepler light curves}

The data set of eclipsing binaries provided in the \kepler Eclipsing Binary Catalog \citep{kepler7} consists of 2876 binary systems, including eclipsing binary and multiple systems, ellipsoidal variables, and eccentric binaries with dynamical distortions, more commonly known as heartbeat stars \citep{thompson2012}. All eclipsing binary light curves have geometrical light curve parameters (eclipse depths, widths and separation) determined with the \polyfit model of \citet{ebaipaper}. Classification of the eclipsing binary systems is done via LLE \citep{kepler3}, which yields a number between 0 and 1 that corresponds to the morphology of the system: 0 - 0.5 values are predominantly assigned to detached systems, 0.5 - 0.7 to semi-detached systems, 0.7 - 0.8 to contact binaries and 0.8 - 1 to ellipsoidal variables, while heartbeat stars do not have an assigned value. This parameter is denoted as the morphology parameter (hereafter {\sl LLE morph}) and is used as a reference for evaluation of our classification methods.

\subsubsection{\gaiasampled \kepler light curves}
\label{Sect:gaiasampled}

To simulate \gaia time-sampling of the light curves we make use of AGISLab \citep{2012A&A...543A..15H}, which is able to predict the transit times of specific sky locations based on the programmed scanning laws. Observation times were computed for a nominal five year \gaia mission using the Nominal Scanning Law \footnote{Alternative scanning laws were used during two months prior to the start of the Nominal Scanning Law (NSL), but those did not intersect with the Kepler field.}, see e.g. sect.~5.2 of \citet{2016A&A...595A...1G}, for the 2876 Kepler Eclipsing binaries at their original sky-positions in the Kepler field.
The resulting number of field-of-view transits is between 69 and 105 with an average of 87 neglecting any (potential) dead-time.

With per-target \gaia timestamps available from the scanning-law expectations, we phased them according to the respective target ephemerides, and we linearly interpolated \kepler light curves at those phases to obtain simulated \gaia flux values. We then unfolded the phases back into time space and used those light curves as pseudo-\gaia observations.

\kepler per-point uncertainties are assigned statistically, based on the crowding metric and catalog magnitude. Considering that all \kepler targets are on the bright \gaia end, the pseudo-\gaia light curves will be overwhelmingly dominated by shot noise \citep{christiansen2012}. Because of that we do not assign any per-point uncertainty to data and assume that intrinsic light curve variability in \kepler data has a global noise level representative of \gaia as well. Thus, our analysis is valid for \gaia targets in the shot-noise regime, while better noise models would be needed for fainter targets (see \citealt{jordi2010}). At this time no such comparison data set is available so we retain a simplified treatment of noise.


\subsection{Fitting models}
\label{Sect:fitting}

\subsubsection{Polyfit}
\label{Sect:fitting_Polyfit}

The \polyfit analytical model is a polynomial chain fit to the data \citep{ebaipaper}. Individual polynomials in the chain are connected at knots, whose placement is determined
iteratively, by minimizing the overall $\chi^2$ value of the fit. The chain is required to be connected and smoothly wrapped in phase space, but not necessarily differentiable at the knots. This way the polynomial fits, or {\sl polyfits}), are able to reproduce the discrete breaks in light curve flux caused by eclipses. The knots are typically positioned at the top of ingress and egress of the primary
and secondary eclipses, each pair spanning a polynomial. For characterizing \kepler light curves, we use four knots and four quadratic polynomials, following \citet{ebaipaper}.

\subsubsection{Two-Gaussian model}
\label{Sect:fitting_TwoGaussian}

The \twog models aim at characterizing the eclipses and tidal-induced ellipsoidal variability of eclipsing binaries using simple mathematical functions that are fitted to their folded light curves (Mowlavi et al, submitted). 
The geometry of each eclipse is modeled with a base Gaussian function $G_{\mu_i,\,d_i,\,\sigma_i}(\varphi)$ of depth $d_i$ and width $\sigma_i$ located at phase $\varphi=\mu_i$, where the index $i$ denotes the primary ($i=1$) and secondary ($i=2$) eclipse.
The base function is mirrored on phase intervals from $\varphi=-2$ to +2 in order to satisfy the boundary conditions of the periodic signal.
The tidal-induced ellipsoidal variability, on the other hand, is modeled with a cosine function with a period equal to half the orbital period.
In order to avoid an overfit of the data with non-significant components, various models with different combinations of these functions are first fitted to the folded light curves.
The models range from a simple constant model to a full two-Gaussian model with ellipsoidal variability.
The model with the highest Bayesian Information Criterion is then retained.

The light curve geometries induced by the eclipses and ellipsoidal variability are in reality more complex than what can be modeled with simple Gaussian and cosine functions, and it is not the aim of the \twog model to provide a precise model of eclipsing binaries.
However, the \twog model can adequately estimate, in the majority of cases, the eclipse widths and depths, inter-eclipse separation, and ellipsoidal-like variability amplitude, all of which are used in this study to classify eclipsing binaries from their light curve geometries.


\subsection{Light curve classification}\label{Sect:classification}

$t$-SNE is a dimensionality reduction algorithm that is steadily gaining popularity in the scientific community due to its capability to overcome the ``crowding problem" present in many other dimensionality reduction techniques (e.g. LLE, SNE, Isomap; \citealt{tsne}) and thus provides a perfect tool for visualizing high-dimensional data based solely on their similarity, without the need to provide additional data attributes. A $t$-SNE visualization of the original \kepler data set is available in \citet{kepler7}. In this study, we extend this qualitative visualization technique with quantitative classification based on DBSCAN.

\subsubsection{$t$-SNE}
      
The $t$-SNE algorithm is a modified version of the Stochastic Neighbor Embedding technique and has a specific appeal for visualizing data, since it is capable of revealing both global and local structure in terms of clustering data with respect to similarity. In practice, it takes only one input parameter that defines the configuration of the output map. The so-called perplexity ($perp$) parameter \citep{tsne} is similar to the number of nearest neighbors in other methods, with the difference that it leaves it up to the method to determine the number of nearest neighbors, based on the data density. This in turn means that the data themselves affect the number of nearest neighbors, which might vary from point to point.

$t$-SNE defines data similarities in terms of conditional probabilities in the high-dimensional data space and their low-dimensional projection. Neighbors of a data point in the high-dimensional data space are picked in proportion to their probability density under a Gaussian, whose variance is determined for each point separately, based on the perplexity value. Therefore, the similarity of two data points is equivalent to a conditional probability. In $t$-SNE, the conditional probability is replaced by a joint probability that depends on the number of data points, which ensures that all data points contribute to the cost function by a significant amount, including the outliers. The conditional probability of the corresponding low-dimensional counterparts in SNE is also defined in terms of a Gaussian probability distribution, but $t$-SNE has introduced a symmetrized Student $t$-distribution. This allows for a higher dispersion of data points in the low-dimensional map and overcomes the crowding problem that results from the overlapping of clusters in the embedding, since a moderate distance in the high-dimensional map can be represented well by larger distances in its low-dimensional counterpart \citep{tsne}. We have chosen values towards the upper recommended values of the perplexity parameter (30--50), which result in maps with well-defined separate clusters than can be efficiently used for visual inspection or quantitative classification. Lower perplexity values produce too many small clusters that do not bear any significant information, while higher values produce embeddings similar to the $perp=50$ value.

\subsubsection{DBSCAN}

The $t$-SNE algorithm does not provide means for classification, but merely visualization of data. However, choosing an appropriate value of the perplexity parameter and implementing a two-step projection ($ND\rightarrow3D\rightarrow2D$) often results in spatially well localized projections of data grouped in several clusters, which allows for the implementation of clustering algorithms to find and isolate different groups. In this step of the analysis we have applied the DBSCAN algorithm \citep{dbscan}, which groups points that are closely packed together (points with many nearby neighbors) and marks as outliers points that lie scattered in low-density regions. The result depends on two parameters: the maximum distance ($\varepsilon$) of all points in the same cluster from a core point and the minimum number of points ($MinPts$) required to form a dense region. It starts with an arbitrary starting point that has not been visited and once that point's $\varepsilon$-neighborhood is retrieved, a cluster is defined if it contains sufficiently many points. Otherwise, the point is labeled as noise, but this point might later be found in a sufficiently sized $\varepsilon$-environment of a different point and be made part of a cluster. The result is then a list of labeled clusters and the objects corresponding to each of them can be readily retrieved and further analyzed.


\section{Results}\label{Sect:Results}

\subsection{Light curve fits}

\subsubsection{\kepler data}
\label{Sect:ModelComparison_Kepler}

We have studied a subset of 2861 eclipsing binaries from the Kepler Eclising Binaries catalog in order to evaluate the performance of the \twog model and the classification with $t$-SNE and DBSCAN. The subset is formed by excluding light curves of higher hierarchy objects in multiple systems. For systems with two ephemerides in the catalog, we use only the light curve with the shortest period, while for systems with three or four ephemerides, we keep the light curves corresponding to the two shortest periods. 

\begin{figure}[t]
   
   \begin{center}
     \includegraphics[width=\hsize]{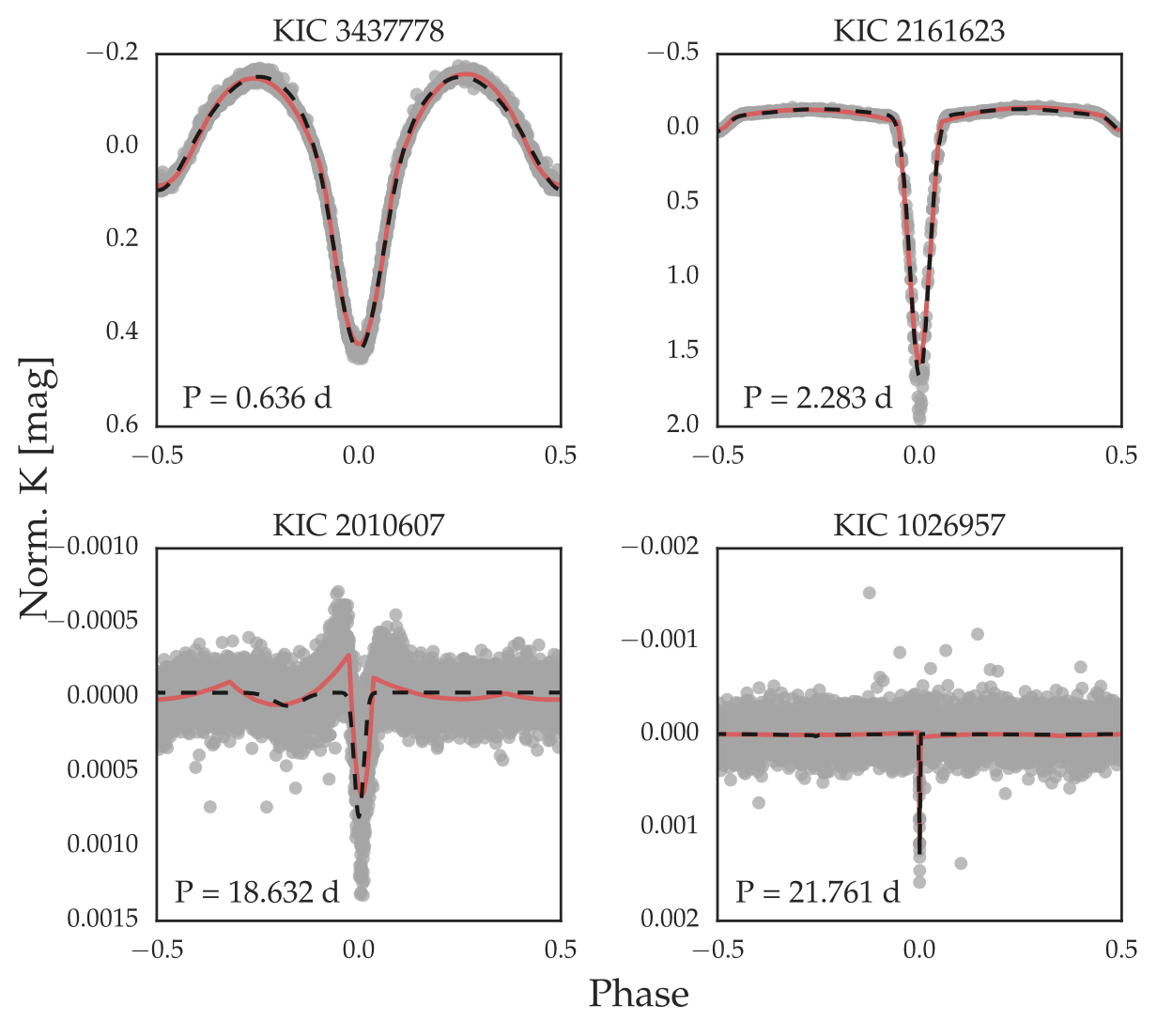}
     \caption{Several examples of \kepler light curves and their respective model fits with {\sl polyfits} and {\sl {\sl two-Gaussians}}. The plots show the observed \kepler light curve (grey dots) in normalized \kepler ($K$) magnitude, \polyfit model (solid red line) and \twog model (dashed black line). Magnitudes are obtained from the \kepler detrended flux and normalized to a reference value of $0$ out of eclipse.}
     \label{Fig:lcfits}
    \end{center}
    
\end{figure}

Several examples of \kepler light curves and their models fits are given in Fig.~\ref{Fig:lcfits}. 

\begin{table}[t]
\caption{The rate of eclipse identification by the two models on the set of 2861 phase-folded \kepler and \gaiasampled light curves.}
\label{Tab:identification-rate1}

\begin{tabular}{l c c}
\hline
\hline
	& Primary	& Secondary \\
Model (\kepler data set)	& eclipse [\%]	& eclipse [\%]	\\ 
\hline
{\sl Polyfit} and \twog 	& 95.3		& 66.8		\\
None	& 0.0		& 14.1		\\
Only {\sl polyfits}	& 4.7		& 3.7		\\
Only {\sl two-Gaussians}	& 0.0		& 15.4		\\
\hline
Data set (\twog model) & 		& 		\\
\hline
\kepler and \gaiasampled	& 63.8           & 50.0	\\
None	& 4.7               & 16.5		\\
Only \kepler	& 31.4            & 32.2		\\
Only \gaiasampled	& 0.10            & 1.3		\\
\hline
\end{tabular}
\end{table}

\begin{figure}
   \begin{center}
     \includegraphics[width=\hsize]{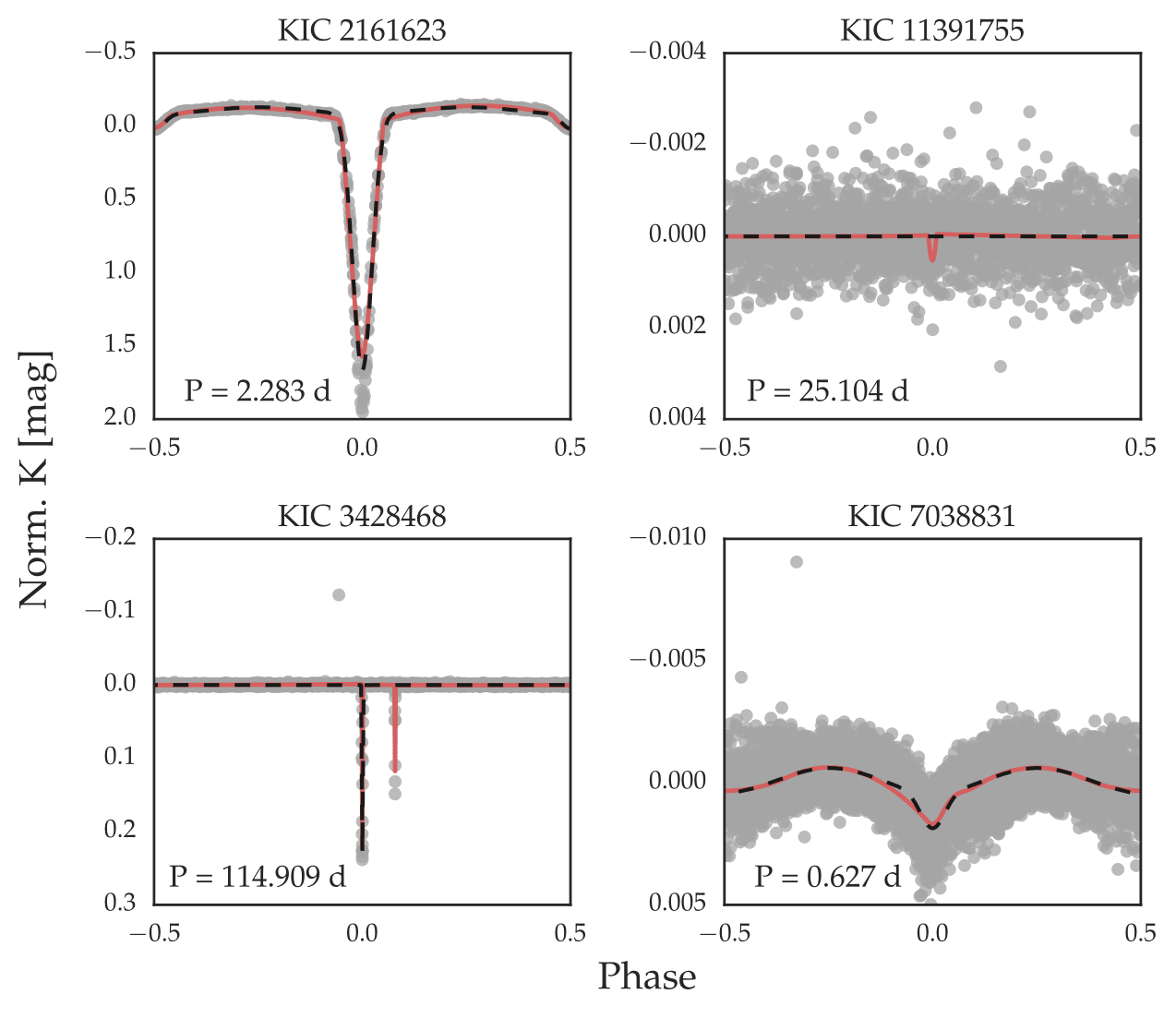}
     \caption{Examples of \kepler light curves where {\sl polyfits} and {\sl two-Gaussians} give discrepant results. The plots show the observed \kepler light curve (grey dots) in normalized \kepler ($K$) magnitude, \polyfit model (solid red line) and \twog model (dashed black line). Top left: a light curve where both models agree; top right: no primary eclipse detected by the \twog model; bottom left: no secondary eclipse detected by the \twog model; bottom right: no secondary eclipse detected by \polyfit.}
     \label{fig:discrepant_lcs}
    \end{center}
\end{figure}

The eclipse detection overlap between the two models is given in Table~\ref{Tab:identification-rate1}. The \twog model fails to detect a primary eclipse for approximately 5\% of the light curves. They are shown to correspond to very noise light curves, where an actual eclipse is practically invisible, even if present. The primary eclipse detected by {\sl polyfits} in these systems is negligible, and we conclude that the failure of the \twog model to detect it is not due to deficiencies of the model, but is governed by the data themselves. There is also a small portion (3.7\%) of the light curves where the \twog model fails to detect a secondary eclipse. These are mainly attributed to very narrow secondary eclipses, that do not contain many data points, as well as eccentric binaries, where the initial conditions are not suitable for the models to converge to a Gaussian at the position of the secondary eclipse. A relatively large number of cases ($15.4\%$) have an identified secondary eclipse by the \twog model only, where a secondary Gaussian function has been fitted to the inter-eclipse variability. The differences in the model results are thus mainly a consequence of how each model adapts to the particular structure of the data. Examples of light curves that illustrate these discrepancies are given in Fig.~\ref{fig:discrepant_lcs}.

For many of the cases the \polyfit and \twog models look almost identical, but the underlying model parameters can differ greatly, in particular the eclipse widths. {\sl Polyfits} compute them based on the positions of the knots in the polynomial chain \citep{ebaipaper}, while the eclipse width in the \twog model corresponds to the widths of the Gaussian functions at a reference magnitude equal to 2\% of the respective depth. For contact and ellipsoidal systems, which show continuous variability, the polynomial knots are usually positioned midway through the eclipses, resulting in a width of $\Delta\varphi\sim0.25$, while in the \twog model the eclipse width of these objects would typically saturate at $\Delta\varphi\sim0.5$ and is limited to $\Delta\varphi=0.4$. 
This disparity should be taken into account when comparing values from the \twog model and the \polyfit values given in the \kepler Eclipsing Binaries catalog.

\subsubsection{\gaiasampled data}

\begin{figure}[t]
   \begin{center}
     \includegraphics[width=\hsize]{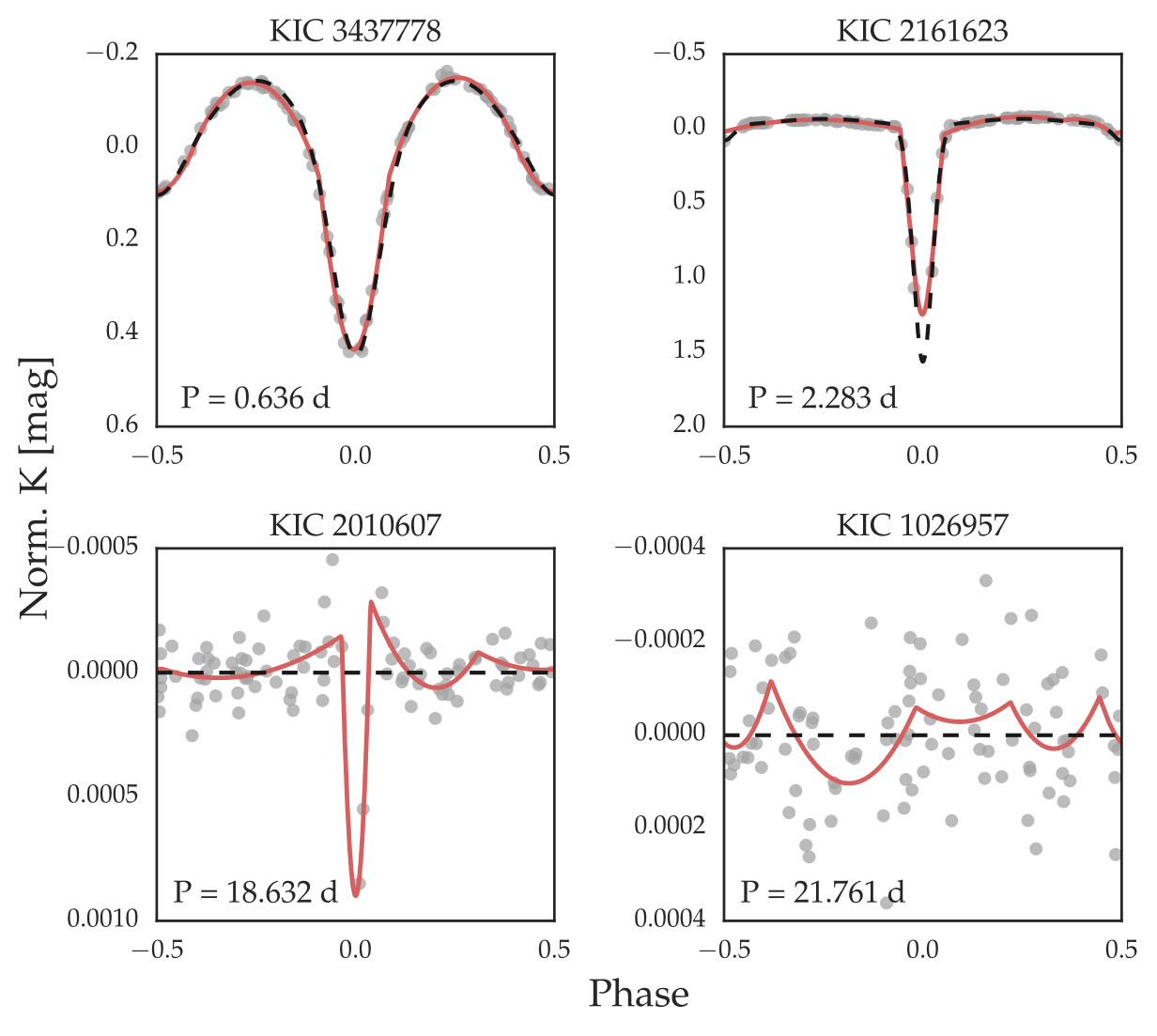}

     \caption{Same as Fig.~\ref{Fig:lcfits}, but for \gaiasampled data. Top left: good quality data and both matching fits. Top right: good data, slightly discrepant fits. Bottom left: bad phase coverage in eclipse, {\sl two-Gaussians} fit a constant model, while {\sl polyfits} find eclipse. Bottom right: bad quality light curve and corresponding model fits.}
          \label{Fig:pf_vs_2g_lcs}
    \end{center}

   \end{figure}

Unlike {\sl Kepler}, \gaia light curves have sparse (irregular) sampling. This introduces an additional difficulty to the models and is especially notable in {\sl polyfits}, which in the absence of a well-defined light curve, tend to fit the polynomial chain to the out-of-eclipse noise. The \twog models, on the other hand, will either fit a constant model to the noise or produce a smooth curve resembling a physical light curve. This is a great advantage when dealing with light curves with well-defined eclipses, but noisy inter-eclipses, as well as for discarding cases where the points in an eclipse cannot be used for any quantitative analysis (bottom left panel of Fig.~\ref{Fig:pf_vs_2g_lcs}).

The eclipse identification rate with {\sl two-Gaussian} models on the \gaiasampled data set  again shows a loss of about $31.5\%$ of the primary eclipse and an additional $16.8\%$ of secondary eclipse identifications, which is a consequence of the deterioration of light curve quality. This gives an estimate of $\sim32\%$ of Kepler Eclipsing Binaries that will likely not be identified with \gaia sampling. A summary of these results is given in Table~\ref{Tab:identification-rate1}.


\subsection{Light curve classification}

\begin{figure}
\begin{center}
\includegraphics[width=0.8\hsize]{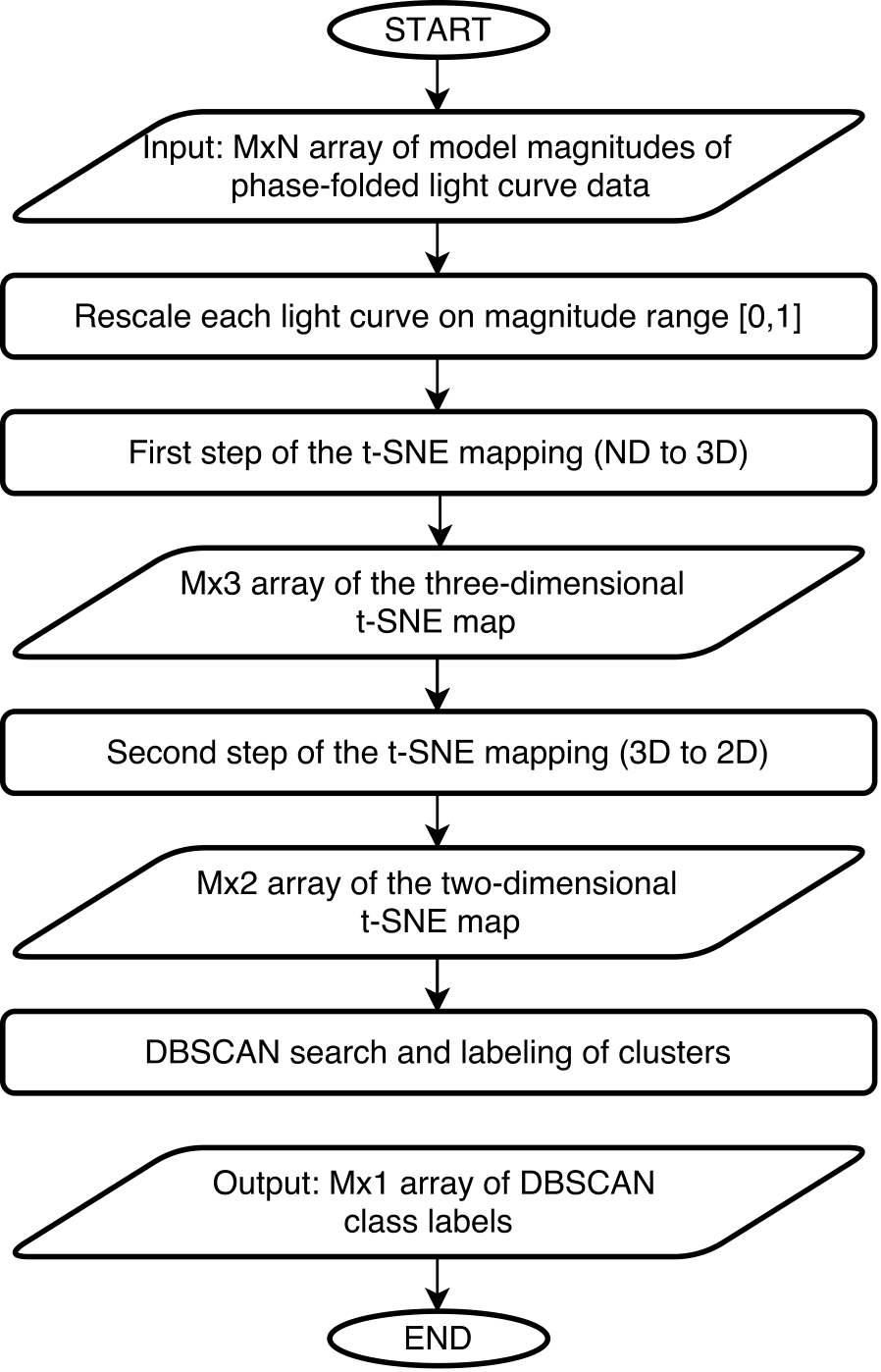}
        \caption{Flowchart of the $t$-SNE+DBSCAN application to a set of $M$ phase-folded light curve models computed in $N$ equidistant phase points.}
        \label{Fig:flowchart}
\end{center}
\end{figure}

We use a combination of the two techniques to classify a data set of $M$ light curves, based on the per-point similarity of the phase-folded light curve models, computed in $N$ phase points. Our \gaiasampled \kepler data set consists of $M=2861$ light curve models computed in $N=1000$ equidistant phase points ranging from $-0.5$ to $0.5$ with imposed periodic boundary conditions.
The input is an $M\times N$ array of the light curve model magnitudes. Each row is first rescaled so that its magnitudes span the range of $[0,1]$, via:

\begin{equation}
    m_i(\mathrm{rescaled}) = \frac{m_i - \min(m_i)}{\max(m_i-\min(m_i))}
\end{equation}
\noindent
where $m_i$ is the array of model magnitudes of the $i$-th source in the input array.
This ensures that the mapping is only sensitive to the light curve shape and unaffected by the different primary eclipse depths. The rescaled magnitude array is first mapped to a three-dimensional map with $t$-SNE. The three-dimensional map serves as input to the second step of the mapping, in which a two-dimensional $t$-SNE map is produced. The two-dimensional $t$-SNE map is then scanned by DBSCAN, which defines a set of clusters and labels them, returning the per-point labels in a $M\times1$ array, which can be then assigned to the original input set of light curves. An illustration of the algorithm flow is given in Fig.~\ref{Fig:flowchart}.

\subsubsection{Polyfit}
\label{Sect:fullDataSet_Polyfits}

\begin{figure*}[t]
      \begin{center}
\includegraphics[width=0.32\hsize]{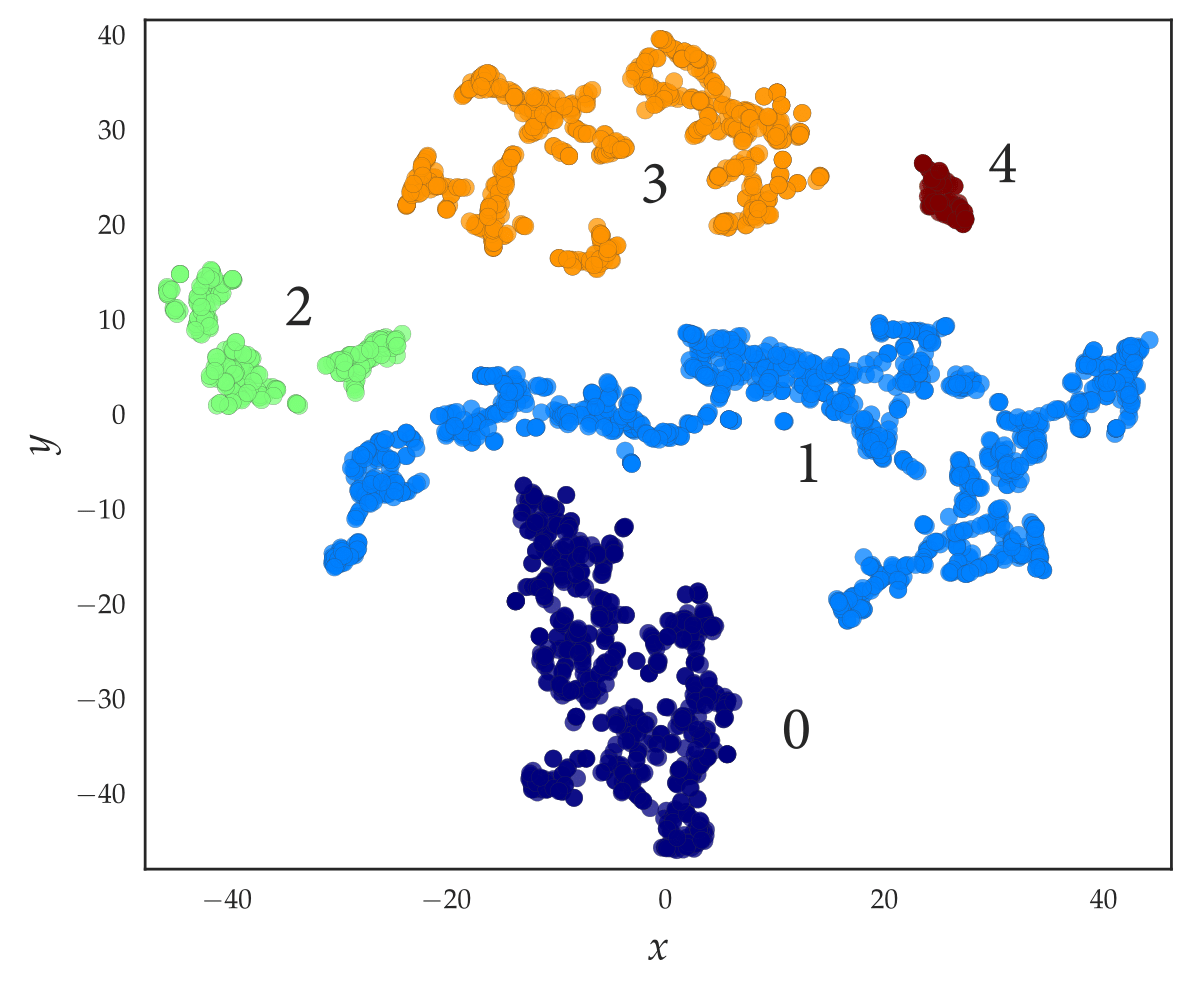}
\includegraphics[width=0.32\hsize]{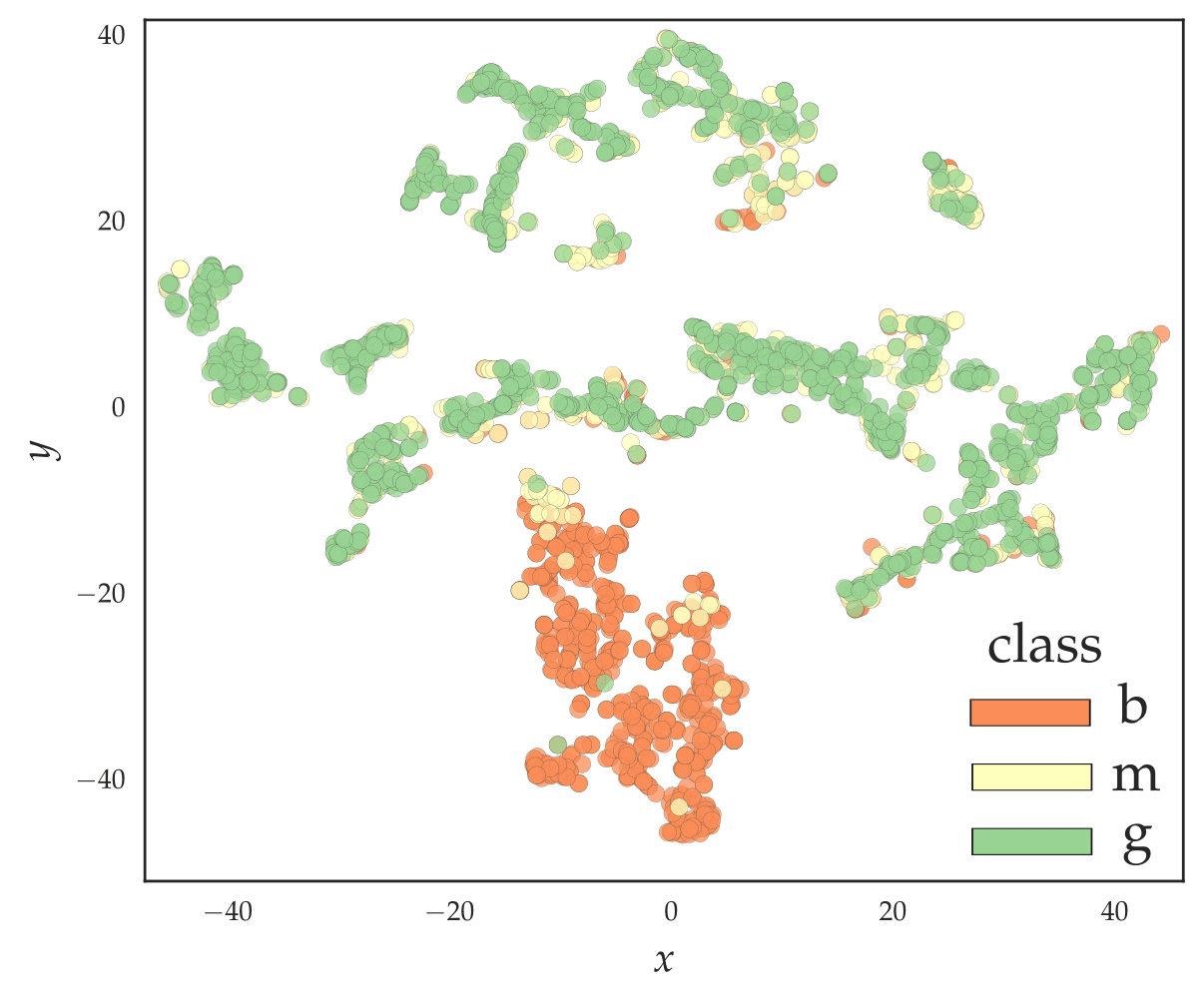}
\includegraphics[width=0.32\hsize]{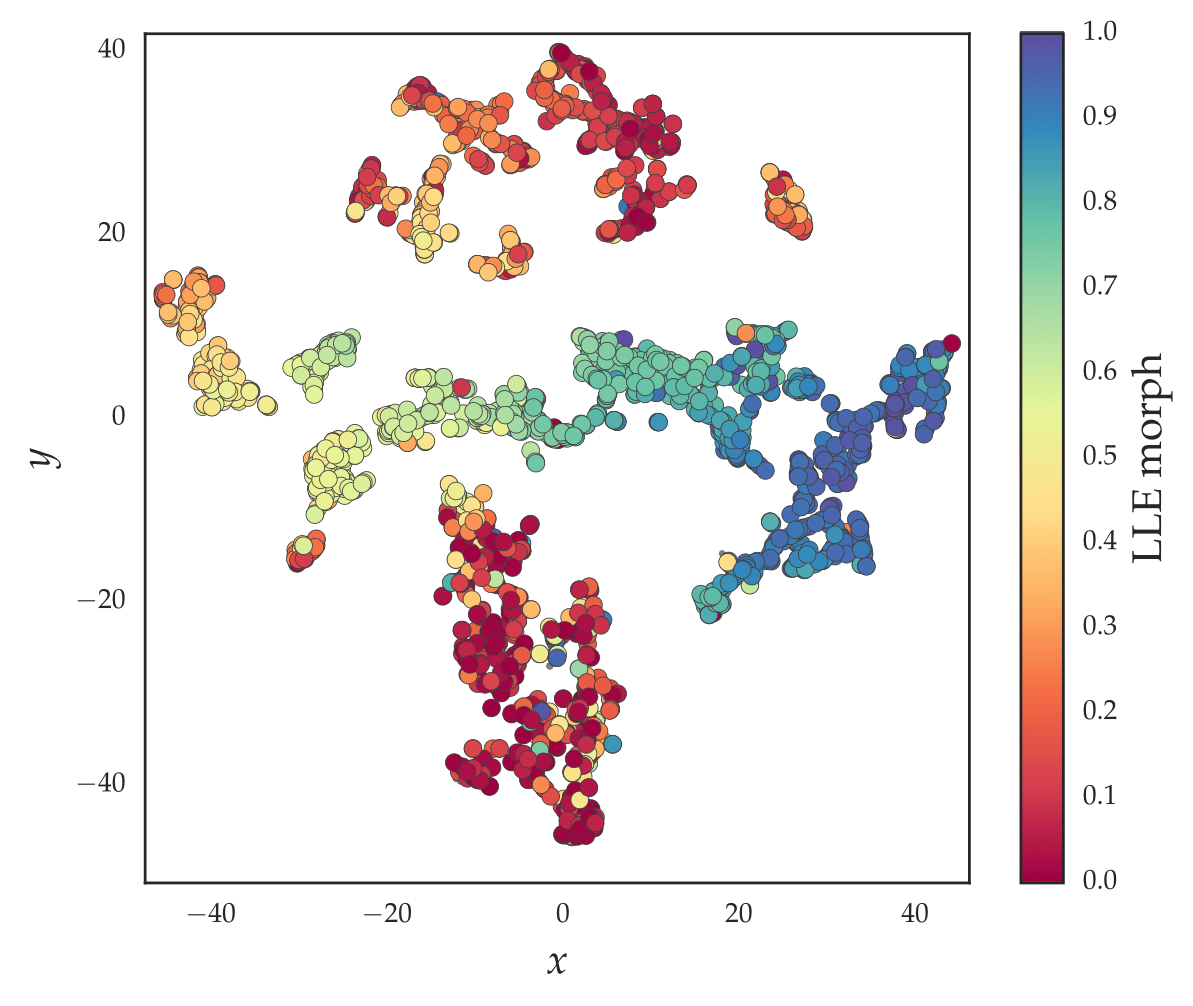}
        \caption{$t$-SNE projection of the full \gaiasampled data set fitted with \polyfit models. Left panel: DBSCAN class flags, middle panel: manual fit quality flags, right panel: \kepler LLE morphology parameter distribution over the $t$-SNE projection.}
        \label{fig:pf_full_tsne}
       \end{center}

\end{figure*}

\begin{figure*}[t]
\centering
\includegraphics[width=0.32\hsize]{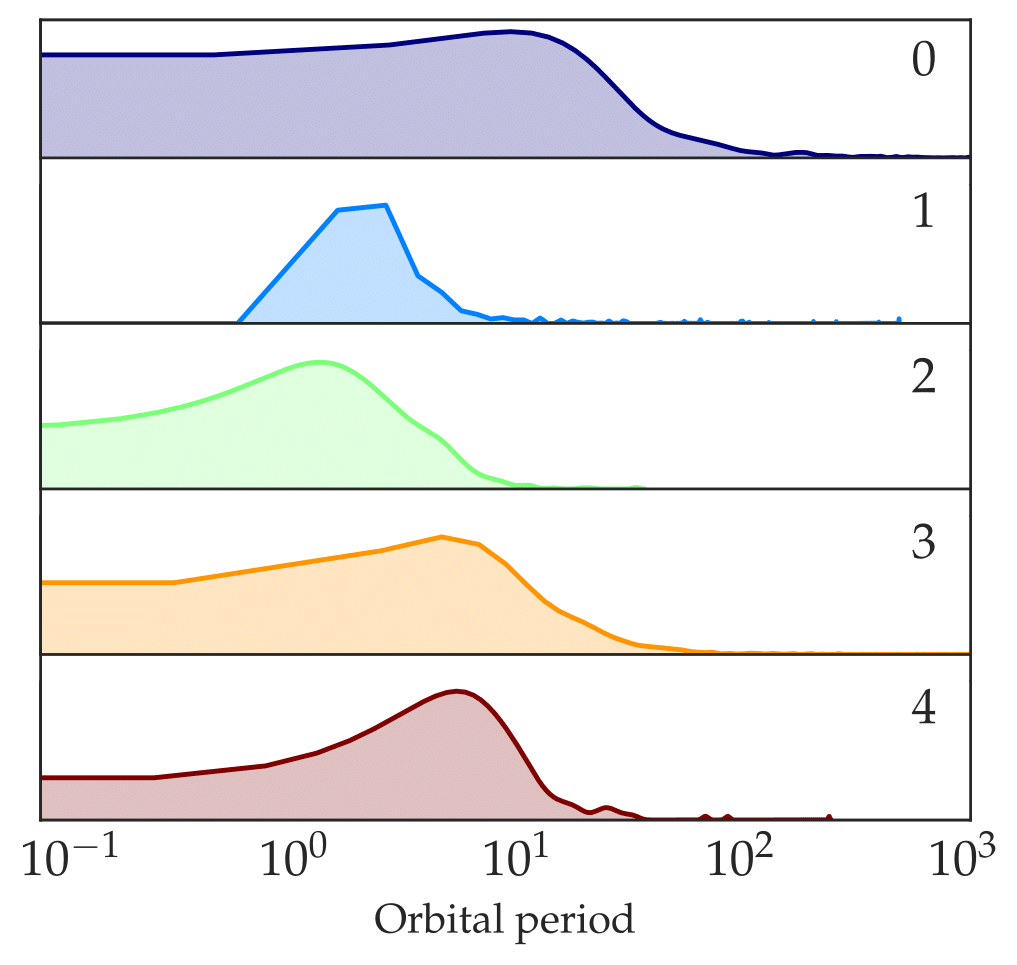}
\includegraphics[width=0.32\hsize]{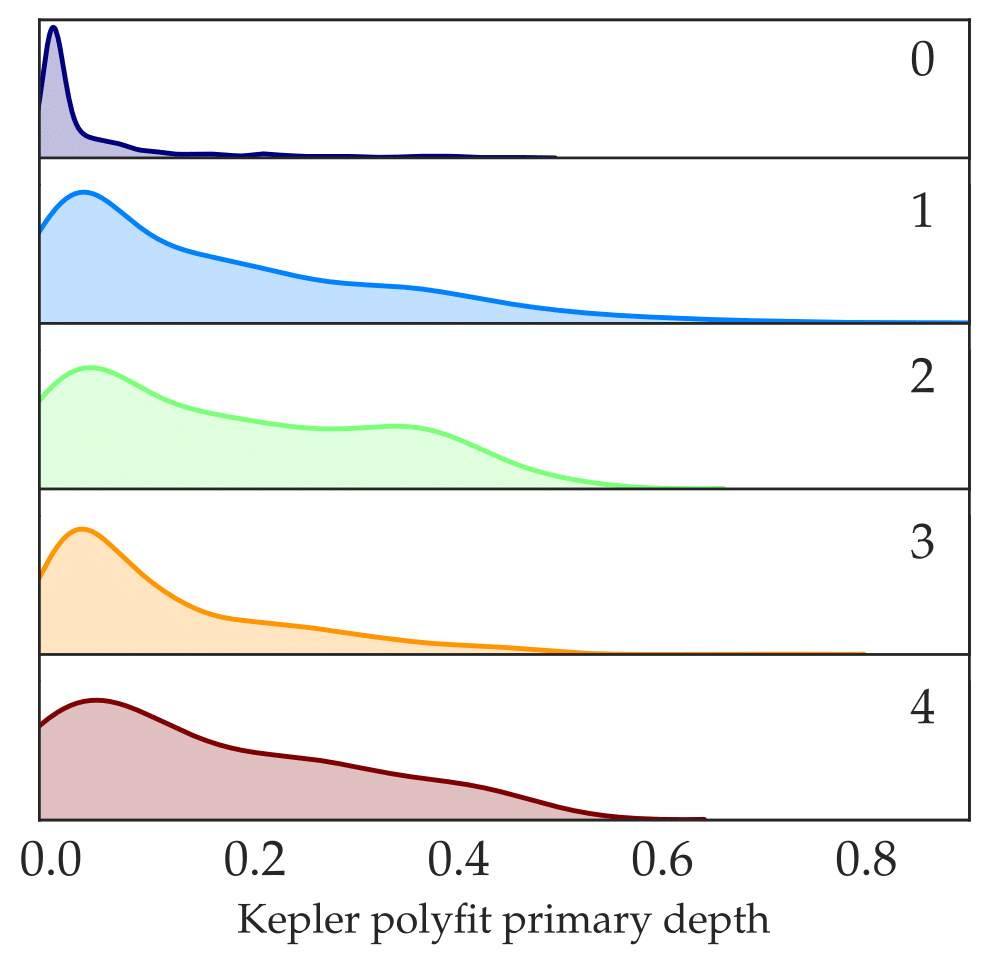}
\includegraphics[width=0.32\hsize]{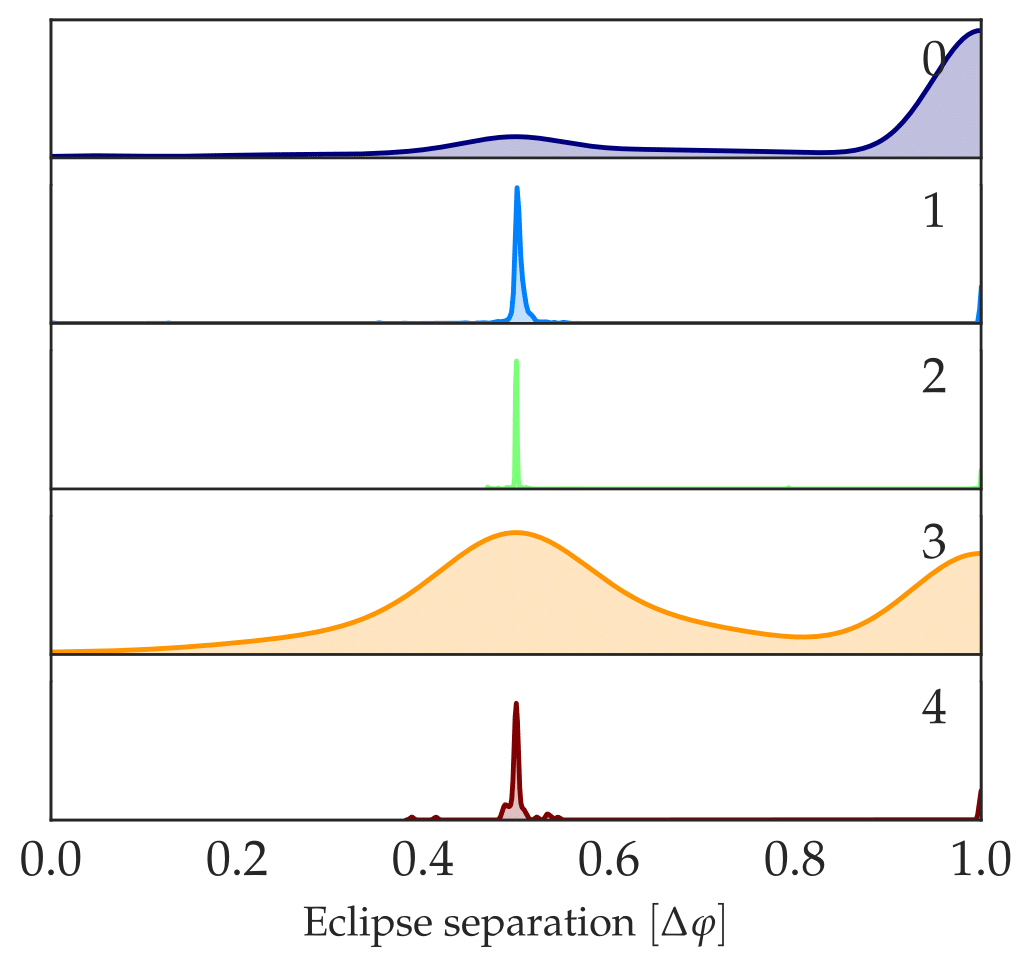}
\caption{Original \kepler value distributions of orbital period, \kepler primary eclipse depth and eclipse separation over the different \polyfit classes. Class descriptions are given in Table~\ref{Tab:dbscan_stats}.}
\label{fig:pfhistograms}
\end{figure*}

We use the Barnes-Hut version of $t$-SNE \citep{VanDerMaaten2014} with $perp_{3D} = 50$ and $perp_{2D} = 45$ and DBSCAN with $\varepsilon = 5$ and $MinPts = 50$ that result in the identification of five \polyfit clusters. The left panel of Fig.~\ref{fig:pf_full_tsne} shows the two-dimensional map color-coded by the detected DBSCAN classes. A visual inspection of the sources pertaining to each of the classes and their true \kepler \polyfit parameters is used to define the descriptive classification provided in Table~\ref{Tab:dbscan_stats}. It shows that the mapping has been driven by the morphology of the systems, ranging from close eclipsing binaries and ellipsoidal variables to detached systems, as well as the relative depth of the eclipses. Phase-folding of the \polyfit models has been performed with the known value of the zero-time reference point of primary eclipse obtained by \kepler light curves, which is why we get a class of sources with a more conspicuous secondary eclipse. Had this value not been provided to the model, classes 3 and 4 would have most likely been merged into one class of light curves with only one conspicuous eclipse, centered at orbital phase $\varphi = 0$. Classes 1 and 2 contain the close binary types (semi-detached --SD--, contact binaries --CB-- and ellipsoidal variables --ELV--), while class 0 contains the poor fits, i.e. {\sl polyfits} which do not resemble a physical light curve.
   
In addition to the automated approach, we have performed a visual evaluation of each \polyfit with three flags: good (g), medium (m) and bad (b) quality fit. Out of the 2861 sources, 1308 (46\%) were marked as good, 770 (27\%) as medium, and only 783 (27\%) as bad. Actually, the DBSCAN class 0 contains $\sim93.6\%$ of the visually flagged poor fits, $\sim 6\%$ of the medium quality and only $\sim 0.4\%$ of good quality fits, which shows that class 0 is predominantly composed of the visually marked bad fits. The distribution of the flags over the $t$-SNE projection is given in the middle panel of Fig.~\ref{fig:pf_full_tsne}. The distribution of the \kepler morphology parameter derived with LLE (\citealt{kepler3}; right panel of Fig.~\ref{fig:pf_full_tsne}) shows the continuous transition from detached ($morph=0$) to contact/ellipsoidal ($morph=1$) binaries over the four classes containing good quality fits, while the poor \polyfit class is shown to be mainly composed of detached systems, whose eclipses are likely not observed in \gaia cadences. 

We use {\sl Kepler}-derived parameters and quantities related to the quality of fits to search for indicators of the underlying causes for the bad data quality resulting in poor fits. The histograms of the orbital period, primary eclipse depth and eclipse separation parameters over the different classes given in Fig.~\ref{fig:pfhistograms} show that class 0 is composed of both long and short period binaries and low to high eclipse depth values, but the low-eclipse depth systems prevail. Based on these parameter distributions, we conclude that class 0 is composed of sources that will likely not show any eclipsing binary features due to the possible miss of eclipse or low eclipse signal buried in the background noise. Classes 3 and 4 show similar parameter distributions, but their light curves have at least one prominent eclipse, thus we can consider those "lucky catch eclipses" that happen in \gaia cadences. The wider distributions of eclipse separation in classes 0, 3 and 4 show that they contain eccentric systems, as well as light curves where the secondary eclipse is not present (eclipse separation equal to 1). This similarity of class 0 to classes 3 and 4 further reinforces the necessity of a filtering approach primarily based on the light curve shape, since no threshold can be set to the values of any of the characteristic light curve parameters that would provide a clear distinction between poor and good quality fits.

\begin{table}[t]
\caption{DBSCAN results statistics for the \polyfit models of the \gaiasampled data set.}

\resizebox{\hsize}{!}{
\begin{tabular}{llll}
\hline\hline
class & \# of sources & \% & class description                                                          \\ \hline
0     & 693                        & 24                         & bad \polyfit \\
1     & 1197                       & 42                          & short-period EBs (SD,CB) and ELVs \\
2     & 251                       & 9                         & short-period EBs (D,SD,CB)                                                      \\
3     & 642                        & 22                         & detached, primary eclipse more conspicuous                \\
4     & 78                         & 3                          & detached, secondary eclipse more conspicous      \\
\hline
\end{tabular}
}

\label{Tab:dbscan_stats}
\end{table}


\subsubsection{Two-Gausssian}
\label{Sect:fullDataSet_TwoGaussian}

\begin{figure}
   \begin{center}
     \includegraphics[width=\hsize]{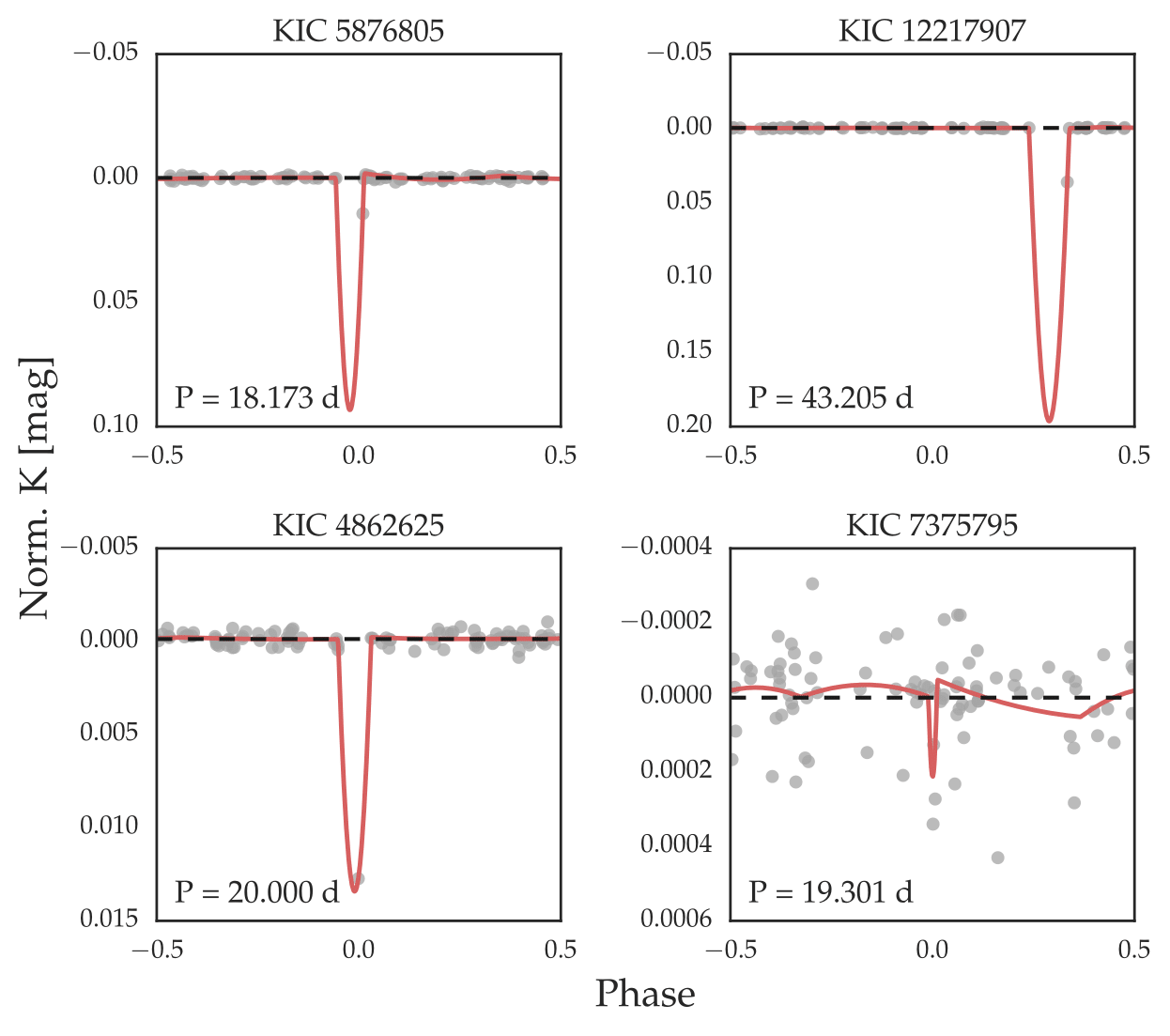}
     \caption{Examples of \gaiasampled light curves where \polyfit fits an eclipse, while {\sl two-Gaussians} do not. The plots show the observed \kepler light curve (grey dots) in normalized \kepler ($K$) magnitude, \polyfit model (solid red line) and \twog model (dashed black line).}
     \label{fig:gs_c_goodpf}
    \end{center}
\end{figure}

The \twog model has a built-in mechanism of identifying the light curves which do not show any eclipses, through the choise of a constant as the best-fit model. Out of the 2861 \gaiasampled light curves, 859 or about 30\% were fitted with a constant model. 575 or 66\% of these light curves also belong to the bad polyfit class, while the remaining 34\% primarily correspond to light curves with one or just a few data-points in eclipses, which the \twog model flags as insignificant, while \polyfit still fits an eclipse. Examples of these light curves are given in Fig.~\ref{fig:gs_c_goodpf}. 

\subsubsection{Filtered data set}

\begin{figure*}[t]
\centering
\includegraphics[width=0.48\hsize]{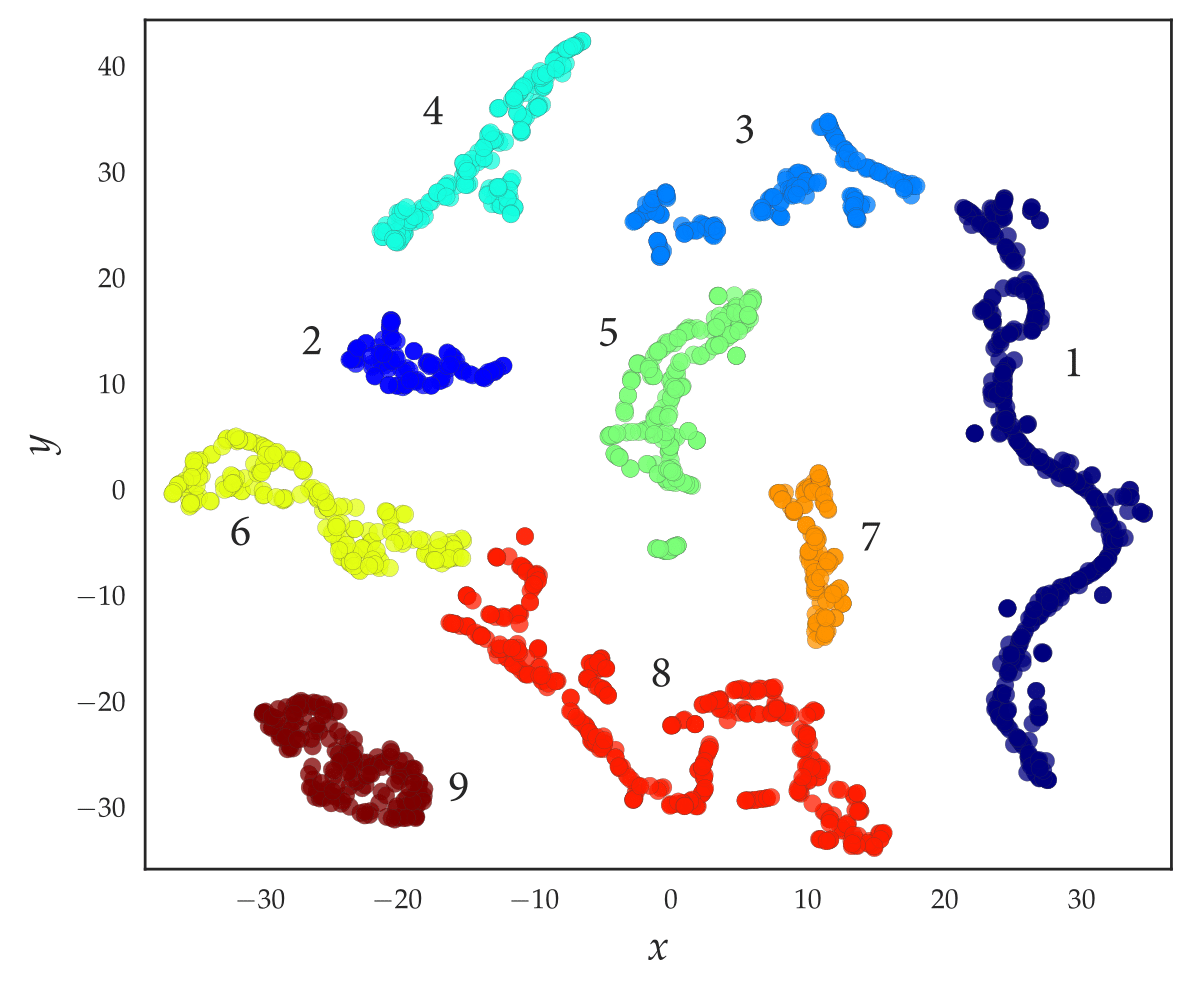}
\includegraphics[width=0.42\hsize]{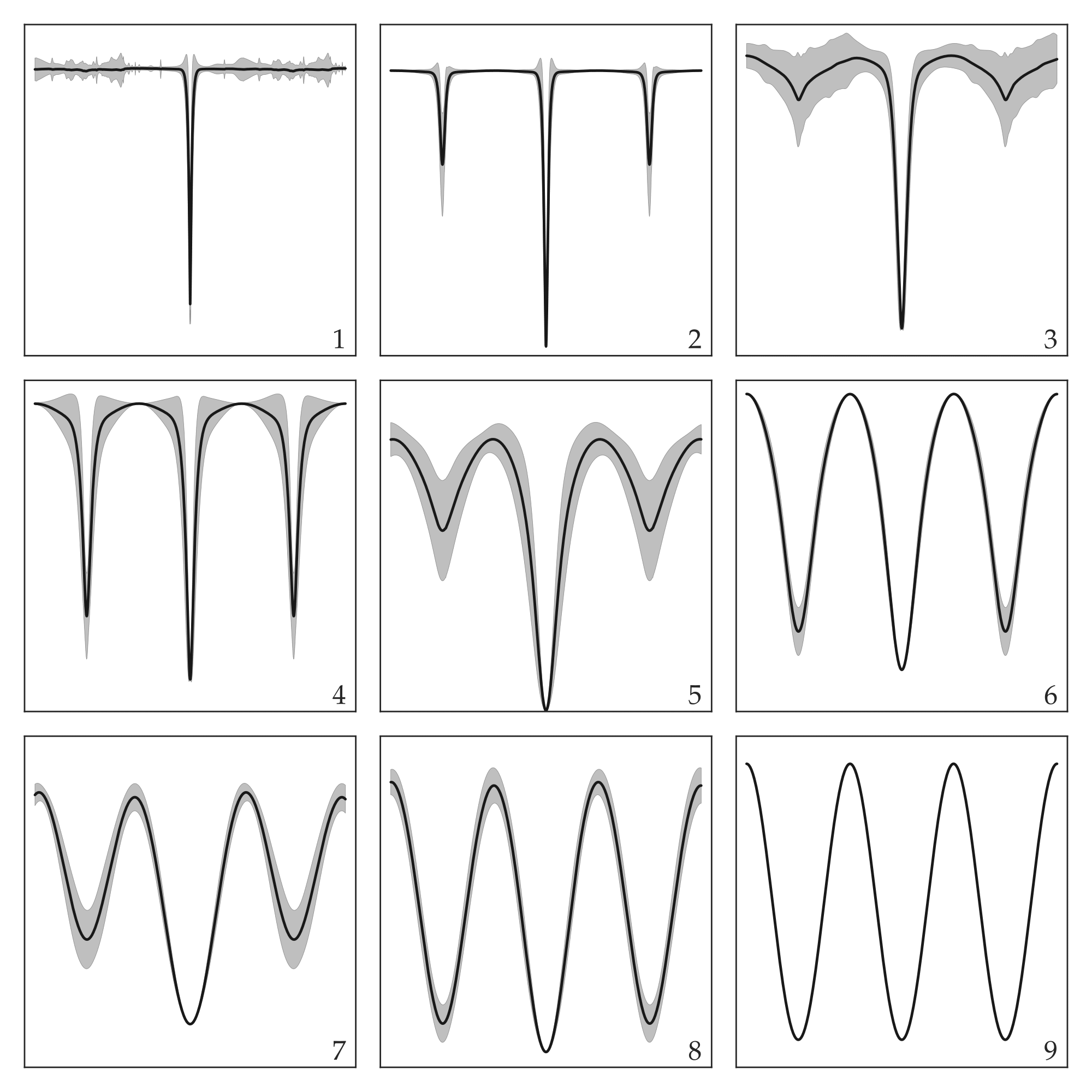}
    \caption{Left panel: $t$-SNE+DBSCAN of filtered \gaiasampled \kepler data set fitted with {\sl two-Gaussians}. Right panel: mean of all normalized light curves in each DBSCAN class, grey shading indicates the region $[-\sigma,\sigma]$ around the computed mean at a given orbital phase.}
     \label{fig:2g_tsne}
   \end{figure*}

The \twog model has a more thorough and consistent approach towards filtering data that is unlikely to pass \gaias eclipsing binary detection pipeline, thus we remove the 859 light curves fitted with the constant model and propose a classification scheme on the remaining 2002 sources.
The application of $t$-SNE+DBSCAN with $perp_{3D}=35$, $perp_{2D}=35$, $\varepsilon=2.6$ and $MinPts=18$ on the \twog models results in nine classes (Fig.~\ref{fig:2g_tsne}), marked from 1 to 9, which can be used to define a classification scheme based on the morphology of the systems and geometry of the light curve fits, given in Table~\ref{Tab:2gfiltered}. Representative light curves of each class are given in Fig.~\ref{fig:2g_tsne}.

\begin{figure*}[t]
\centering
\includegraphics[width=0.45\hsize]{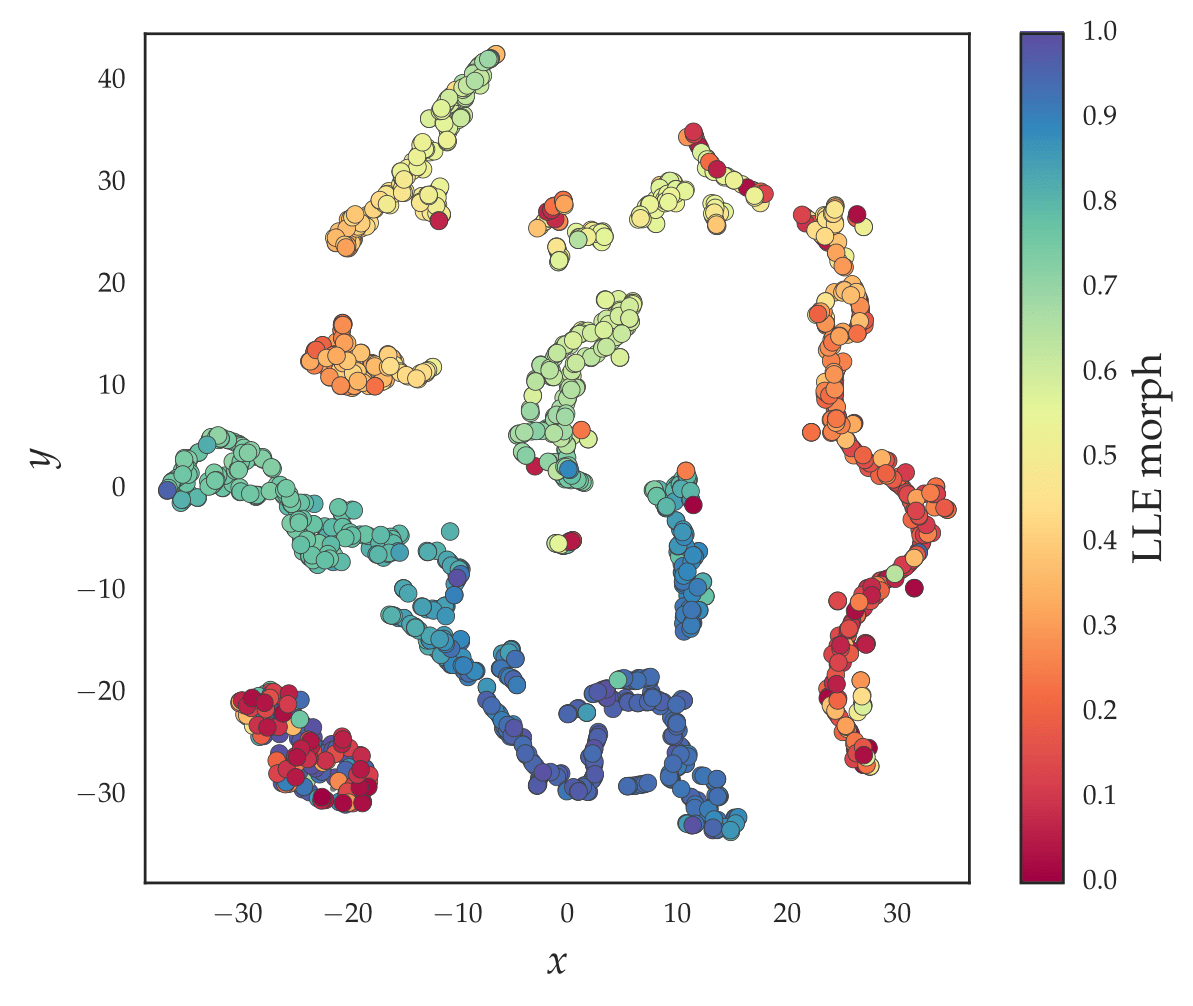}
\includegraphics[width=0.45\hsize]{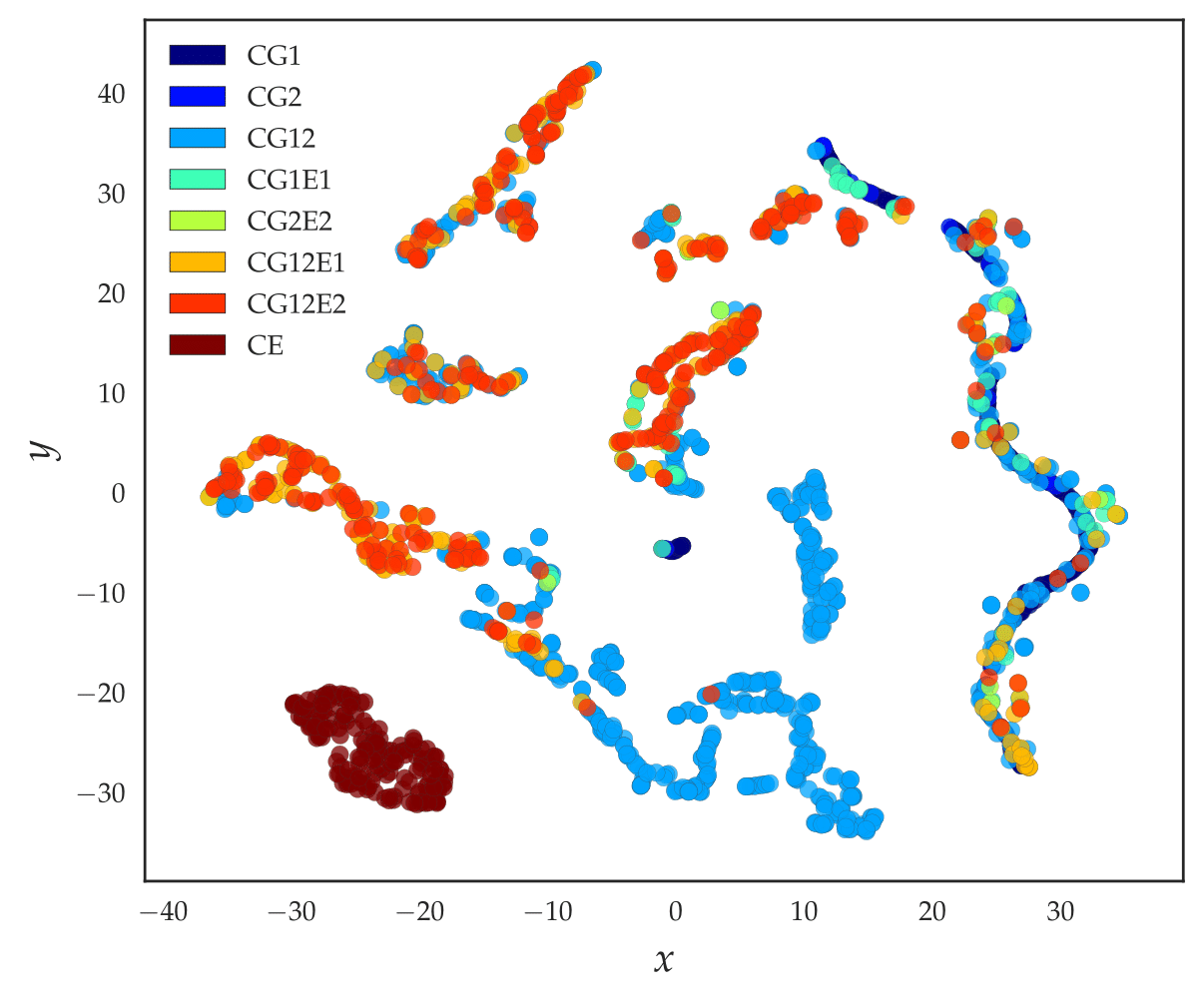}
    \caption{Distributions of the LLE morphology parameter of Kepler EBs (left panel) and \twog models chosen to fit the light curves (right panel) over the $t$-SNE projection of the filtered \gaiasampled \kepler data set fitted with {\sl two-Gaussians}.}
     \label{fig:2g_tsne_dist}
   \end{figure*}

\begin{table*}[t]
\centering

\caption{Proposed classification scheme for the \twog fits on \gaiasampled \kepler data.}
\label{Tab:2gfiltered}
\begin{tabular}{@{}lllclcl@{}}
\hline
\hline
\smallskip
class & class description & subclass & class \# & subclass description & \# of sources & \% \\ \hline
\multirow{2}{*}{D} & \multirow{2}{*}{detached binaries} & D-1 & 1 & one conspicuous eclipse & 429 & 21 \\
 &  & D-2 & 2 & two conspicuous eclipses & 105 & 5 \\
\multirow{2}{*}{D+SD} & \multirow{2}{*}{detached and semi-detached} & D+SD-1 & 3 & one conspicuous eclipse & 178 & 9 \\
 &  & D+SD-2 & 4 & two conspicuous eclipses & 180 & 9 \\
SD+CB & semi-detached and contact binaries & / & 5 &  & 208 & 10 \\
CB & contact binaries & / & 6 &  & 115 & 6 \\
\multirow{2}{*}{CB+ELV} & \multirow{2}{*}{contact binaries and ellipsoidal variables} & CB+ELV-a & 7 & eclipse depth ratio $< 1$ & 212 & 11 \\
 &  & CB+ELV-b & 8 & eclipse depth ratio $\sim 1$ & 403 & 20 \\
ELF & ellipsoidal fits & / & 9 &  & 172 & 9 \\ \hline
\end{tabular}
\end{table*}

\begin{figure*}[t]
\centering
\includegraphics[width=0.32\hsize]{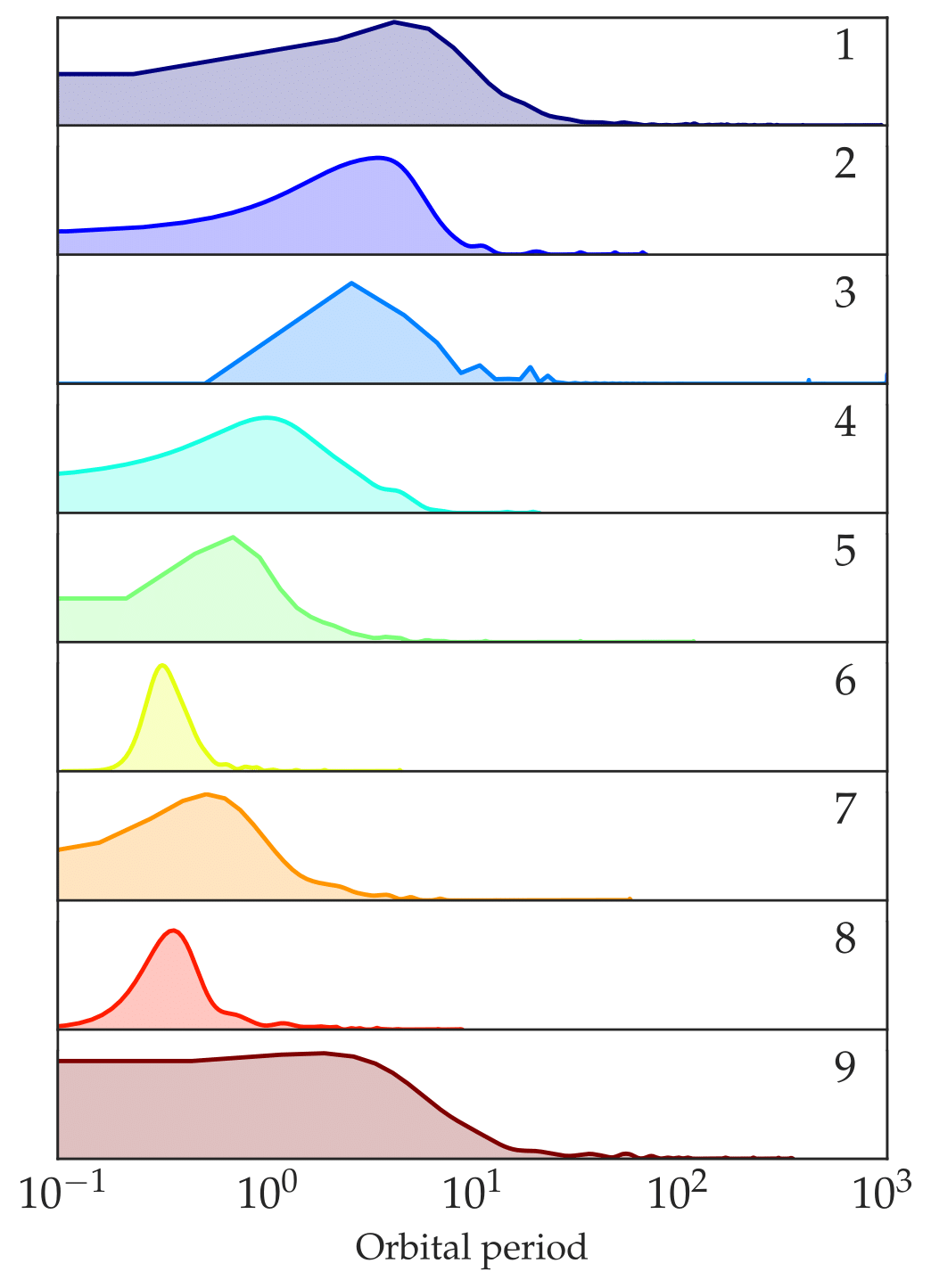}
\includegraphics[width=0.32\hsize]{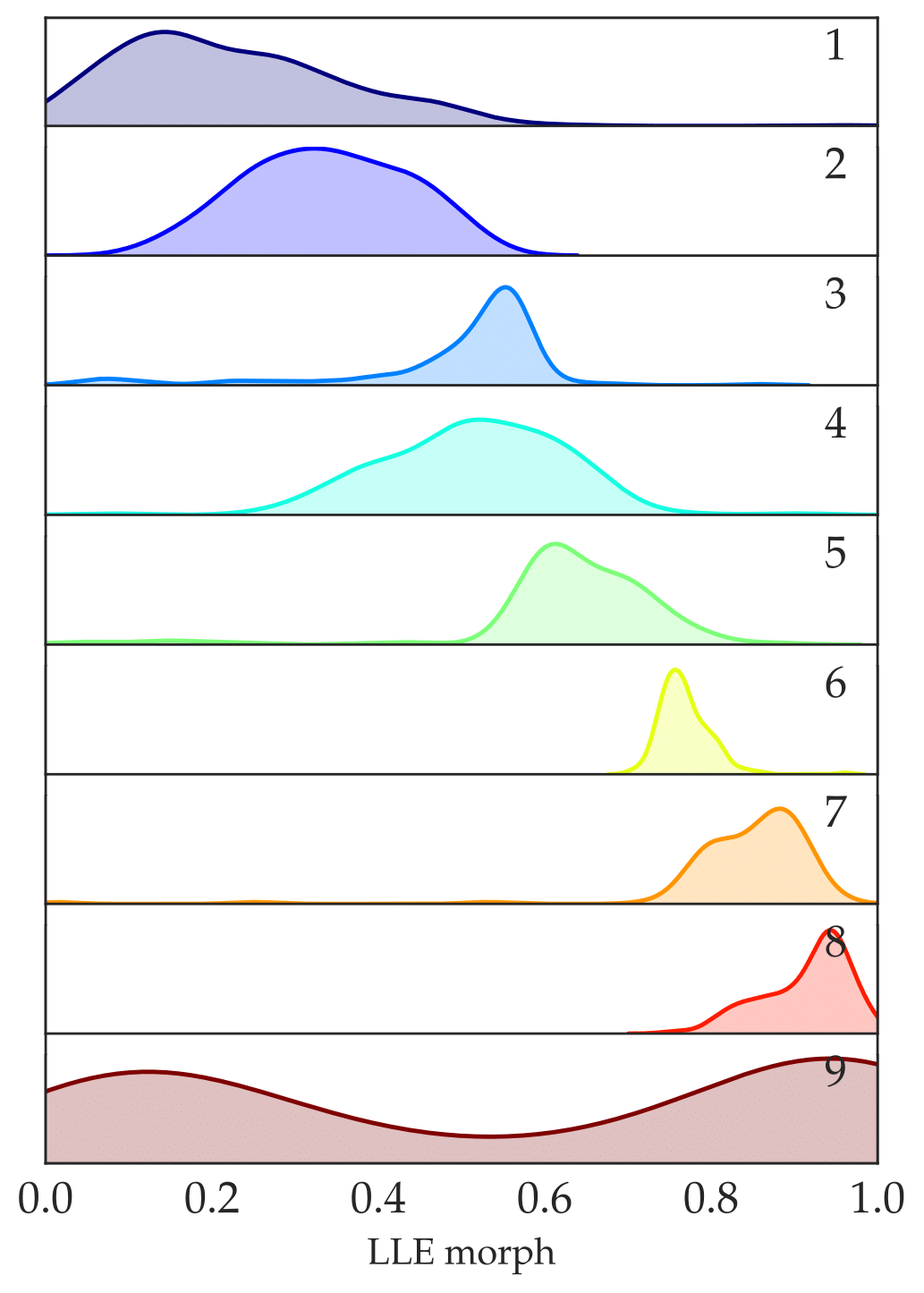}
\includegraphics[width=0.32\hsize]{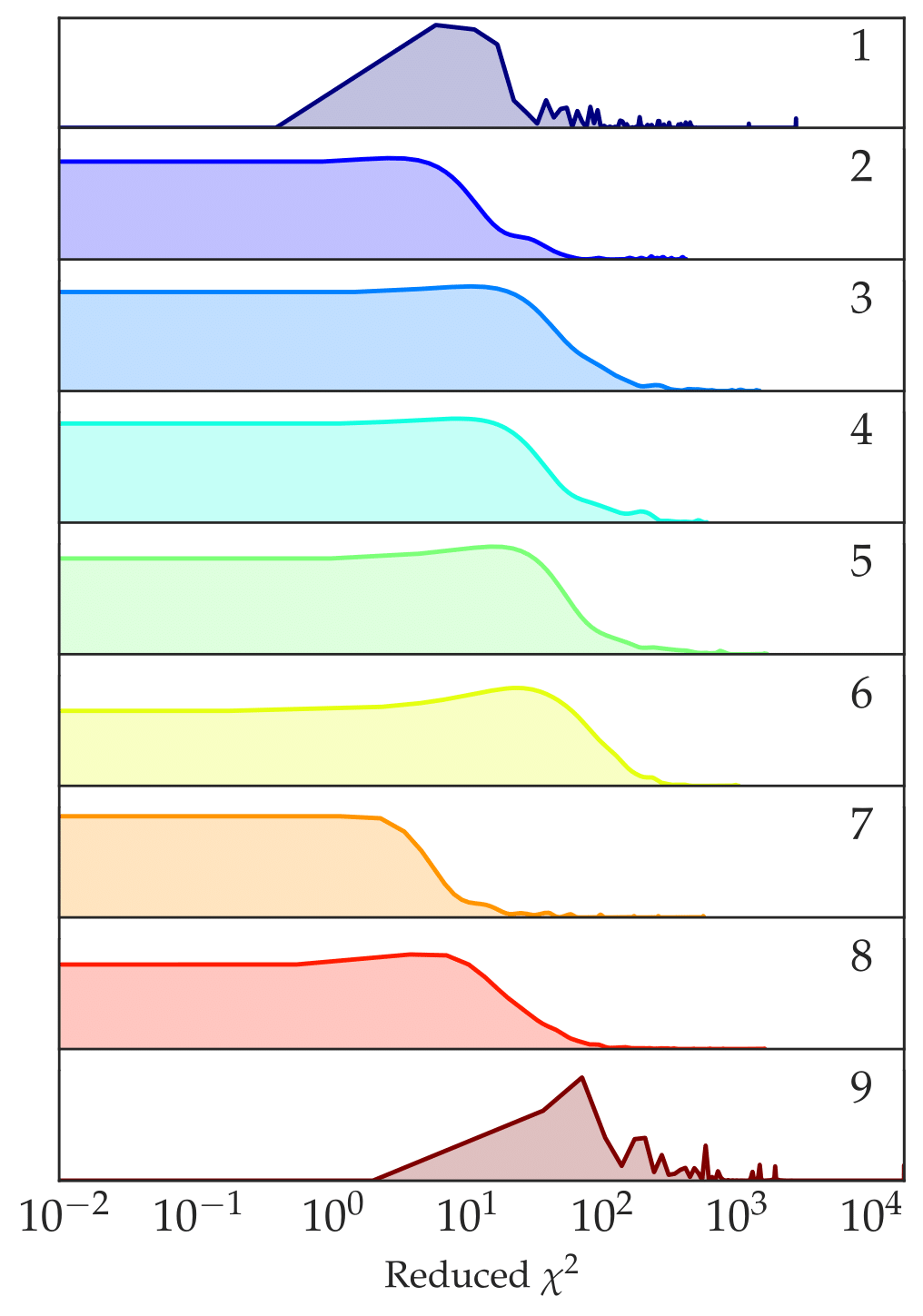}
\caption{Orbital period, LLE morphology parameter and reduced $\chi^2$ distributions over different classes of the projection of the filtered \twog data set.}
\label{fig:2gshistograms}
\end{figure*}

Six main classes have been defined based on the light curve morphology, ranging from detached (D), detached and semi-detached (D+SD), semi-detached and contact binaries (SD+CB), contact binaries (CB), contact binaries and ellipsoidal variables (CB+ELV) and ellipsoidal fits (ELF). A clear distinction between overlapping morphological types among the different classes (D+SD, SD+CB and CB+ELV) cannot be made because their light curve fits are intrinsically similar and the continuous transition from one morphological type to another is an inherent property of the light curve shapes. A more detailed distinction can be made through the inspection of the individual light curve properties, while in some cases full modeling of the system might be required for the accurate determination of its morphological type. The proposed morphological classes are thus an initial indication of the system morphology based on its light curve fit with the \twog model.

The subclasses of each class are based on the geometrical properties of the light curves. The presence and visibility of eclipses define the subclasses in the detached and semi-detached classes: one conspicuous eclipse (D-1 and D+SD-1) or two conspicuous eclipses (D-2 and D+SD-2), while in the CB+ELV subclasses two eclipses are visible in most of the cases, thus the sub-classification is driven by the eclipse or ellipsoidal variation widths and depths. In contact systems, the eclipse widths and depths can also be used as indicators of the physical system parameters, like filling factor and temperature equilibrium. Wider eclipses, characteristic for classes 7 and 8, point to a larger filling factor, while similar eclipse depths, characteristic for classes 6 and 8, point to a system close to thermal equilibrium (e.g. a W UMa star).

The distribution of the \kepler LLE morphology parameter over classes (left panel of Fig.~\ref{fig:2g_tsne_dist} and middle panel of Fig. ~\ref{fig:2gshistograms}), suggests that the proposed classification scheme corresponds to the morphological type of the observed sources determined on the true \kepler light curves. The orbital period distribution (left panel of Fig.~\ref{fig:2gshistograms}) further supports this notion, with transitions from long-periods in the detached to shorter in semi-detached and contact classes. 

The distribution of the different \twog model types over the projection (right panel of Fig.~\ref{fig:2g_tsne_dist}) shows that the projection is highly driven by the choice of the fitting model. This is expected since the model defines the light curve geometry, but the different widths and depths of the eclipses lead to mixing of the models in most of the classes, with the exception of class 9, which is composed solely of ellipsoidal fits.

The distribution of the reduced $\chi^2$ value (right panel of Fig.~\ref{fig:2gshistograms}) is an indicator of the fit quality, which indirectly influences the classification. The classes corresponding to detached systems have both low and high reduced $\chi^2$, due to the small width and contribution of the eclipses to the overall light curve, while as we move towards closer systems with wider and more significant eclipses, the distributions are dominated by lower reduced $\chi^2$ values. This is a valuable indicator of the reliability of light curve parameters provided for each class. 

Class 9 shows the most peculiar parameter distributions of all, pertaining to both very low and very high values of the morphology parameter, as well as predominantly higher reduced $\chi^2$ values. These indicate that class 9 is not only composed of ellipsoidal fits which correspond to true ellipsoidal variables, but also of detached systems where an eclipse has not been observed and the cosine function is fitted to the inter-eclipse scatter. This flags class 9 and the model parameters derived for the sources in it as unreliable and subject to further filtering. 

\section{Conclusions and future prospects}\label{Sect:Conclusions}

In this paper, we have presented for the first time a proposed method of automated reduction and classification of \gaia eclipsing binary data.

Results from both the analysis of the bad cases identified by the {\sl polyfits} and the \twog models, and the comparison of eclipse detection rates of the \twog model applied to real \kepler and \gaiasampled \kepler data, show that about 68\% of all eclipsing binaries in the magnitude interval of \kepler are detectable by \gaia over a five year mission.
The orbital parameters and morphologies derived from \kepler data show that the 32\% non-detectable sources are mainly long-period, detached binaries, with very narrow eclipses that can easily be missed in the $\sim87$ light curve points expected to be observed (on average) in Cygnus by \gaia over the five years. Given that the all-sky average number of observations for \gaia is $\sim67$, we expect the all-sky detectability to be lower than 68\%.
The other type of sources less likely to be detected comprises systems with very low eclipse depths that can easily be buried in the data noise.

We investigated the efficiency of the combined use of the $t$-SNE and DBSCAN algorithms to classify eclipsing binary light curves.
The application to \kepler eclipsing binary light curves sampled at observation times predicted by the \gaia scanning law and characterized with the \twog model shows that the method is successful in identifying six broad classes corresponding to the system morphology (detached, detached+semi-detached, semi-detached+contact binaries, contact binaries, contact binaries+ellipsoidal variables and ellipsoidal fits).

An additional sub-classification is introduced based on the properties of the fitted models (presence and visibility of eclipses and eclipse widths) to distinguish them from the physical properties of the observed systems, which may not be correctly evaluated due to the irregular sampling in \gaia observations (for example, systems with only one observed eclipse may in reality also have a prominent secondary eclipse that was not observed with {\sl Gaia}'s scanning law).

The thorough testing, formulation and implementation of automated reduction and classification techniques are of highest priority in the \gaia era. One aspect of the method that is still open for adjustment on real \gaia data is its performance on data sets of much larger scale than the one used in this study. Several optimization approaches using parametric mapping between the high- and low-dimensional space (i.e. parameteric $t$-SNE) are being considered, as well as a combination of more than one classification approach (see \citealt{suveges2016}). The final aim of this collaborative effort is to provide \gaia data archive users with a clean set of geometrical light curve parameters of eclipsing binaries, an estimate of their credibility, and classification types that would make the selection of a desired data subset as effortless and reliable as possible.

\begin{acknowledgements}
We are grateful to our referee Dr. J.~A.~Caballero for his constructive input, which substantially improved the quality and presentation of the paper.

A.~K. gratefully acknowledges the MSE postdoctoral fellowship of the College of Liberal Arts and Sciences at Villanova University.
\end{acknowledgements}

\bibliographystyle{aa}
\bibliography{bibTex}
\end{document}